\documentclass{aa}                           
\usepackage{natbib,twoopt}
\usepackage{hyperref}

\usepackage{amsmath}
\usepackage{graphicx}                      \usepackage[hypcap]{caption}
\usepackage{subcaption}         
\usepackage{txfonts}
\begin{document}

   \title{DCO$^+$, DCN and N$_2$D$^+$ reveal three different deuteration regimes in the disk around the Herbig Ae star HD163296}

   \subtitle{ }

   \author{V.N. Salinas \inst{1}
          \and
          M.R. Hogerheijde \inst{1,2}
          \and
          G.S. Mathews \inst{3}
          \and
          K.I. \"Oberg \inst{4}
          \and
          C. Qi \inst{4}
          \and
          J.P. Williams \inst{3}
          \and
          D.J. Wilner \inst{4}
          }

   \institute{Leiden Observatory, Leiden University, PO Box 9513,
     2300 RA, Leiden, The Netherlands\\
              \email{salinas@strw.leidenuniv.nl}
              \and
  Anton Pannekoek Institute for Astronomy, University of Amsterdam, Science Park 904, 1098 XH, Amsterdam, The Netherlands
    	\and
  Institute for Astronomy, University of Hawaii, 2680 Woodlawn Dr., Honolulu, HI 96826, USA
   \and
    Department of Astronomy, Harvard University, Cambridge, MA 02138, USA
             }

   \date{Received ; accepted }

 
  \abstract 
{Deuterium fractionation has been  used to study
the  thermal  history  of  pre-stellar  environments.  Their
formation pathways  trace different regions  of the disk and  may shed
light  into  the disk's  physical  structure,  including locations  of
important features for planetary formation. }
{We aim to  constrain the radial extent of  main deuterated species; we are particularly interested
in spatially characterizing the high  and low temperature pathways for
enhancing deuteration of these species. }  
{We  observed the disk  surrounding the  Herbig Ae star  HD 163296
using ALMA  in Band 6  and obtained resolved spectral imaging data of DCO$^+$  ($J$=3--2), DCN
($J$=3--2) and  N$_2$D$^+$ ($J$=3--2) with synthesized beam sizes of 0$\farcs$53$\times$ 0$\farcs$42, 0$\farcs$53$\times$ 0$\farcs$42 and  0$\farcs$50$\times$ 0$\farcs$39  respectively.  We adopt a  physical model  of the
disk from  the literature and use the  3D radiative transfer code  LIME to
estimate an excitation temperature profile for our detected lines. We model the radial emission profiles of DCO$^+$, DCN and
N$_2$D$^+$,  assuming  their  emission  is  optically  thin,  using  a
parametric model  of their  abundances and our  excitation temperature
estimates.}  
{DCO$^+$ can  be described  by  a three-region model, with constant-abundance rings centered at 70 AU, 150 AU and 260 AU.   The  DCN  radial  profile  peaks at  about  ~60  AU  and
N$_2$D$^+$  is  seen in  a  ring  at  ~160 AU.  Simple models  of  both
molecules using  constant abundances  reproduce the data. Assuming reasonable
 average excitation temperatures for the whole disk, their disk-averaged  column densities
(and deuterium  fractionation ratios) are 1.6--2.6$\times 10^{12}$
cm$^{-2}$  (0.04--0.07),  2.9--5.2$\times 10^{12}$  cm$^{-2}$  ($\sim$0.02)   and 1.6--2.5 $\times    10^{11}$   cm$^{-2}$
(0.34--0.45)  for DCO$^+$,  DCN and  N$_2$D$^+$, respectively.}   
{ Our simple best-fit  models show a correlation between the
radial location  of the  first two  rings in DCO$^+$  and the  DCN and
N$_2$D$^+$ abundance distributions that can be interpreted as the high
and low temperature  deuteration pathways  regimes.  The  origin of the third DCO$^+$ 
ring at  260 AU is unknown  but may be due  to a local decrease  of ultraviolet
opacity allowing the photodesorption of CO or due to thermal desorption of CO
as a  consequence of radial drift  and settlement of dust  grains. The
derived deuterium fractionation values agree  with previous estimates of 0.05 for DCO$^+$/HCO$^+$ and 0.02 for DCN/HCN in HD163296, and 0.3-0.5 for N$_2$D$^+$/N$_2$H$^+$ in AS 209, a T Tauri disk. The high  N$_2$D$^+$/N$_2$H$^+$ confirms N$_2$D$^+$ as
a good candidate for tracing ionization in the cold outer disk.}
   
\authorrunning{V.N. Salinas et al.}
\titlerunning{Different deuteration regimes in the disk around HD163296}
   \maketitle
%

\section{Introduction}
\label{sec:intro} So  far, more than  30 deuterated species  have been
detected   towards  pre-stellar   cores,  and   solar  system   bodies
\citep{Ceccarelli2007,Mumma2011}.    Their  deuterium   fractionation,
usually  higher   than  the   D/H  cosmic  ratio   of  $\sim$10$^{-5}$
\citep{Vidal-Madjar1991}, is used to  infer their thermal history, and
in the  case of solar system  bodies, their location within  the solar
nebula at  the time of  formation.  The amount of 
detections of deuterated species  towards  protoplanetary   disks  is   not  as   high as in
proto-stellar environments.  DCO$^+$ has been detected in both T Tauri
disks
\citep{Guilloteau2006,Oberg2010,Oberg2011,Oberg2015,vanDishoeck2003,Huang2017}
and    in   the    Herbig   Ae    disks   HD163296    and   MWC    480
\citep{Qi2015,Mathews2013,Huang2017}. DCN has been observed towards six different disks by \citet{Huang2017} in addition to being previously observed towards TW Hya \citep{Qi2008} and LkCa15 \citep{Oberg2010}. N$_2$D$^+$ has  only  been
recently  observed  in  the  disk  around the  T  Tauri  star AS  209
\citep{Huang2015}.

 The  massive (0.089  M$_\odot$) and
extended  gas-rich disk  around HD163296  is inclined at $44^{\deg}$. Its
proximity  (122 pc)  makes it  an  excellent laboratory  to study  the
spatial   location   of   the   different   deuteration   regimes   in
protoplanetary   disks  \citep{Qi2011,Mathews2013,Perryman1997}. DCO$^+$ emission  was first  seen in  a ring  towards the  disk around
HD163296   and   suggested  as a tracer   of   the  CO   snowline   by
\citet{Mathews2013}. It was later  observed that the emission extended
further inwards  past the CO  snowline at  90 AU traced  by N$_2$H$^+$
\citep{Qi2015}. ALMA cycle 2 observations towards HD163296 of DCO$^+$,
DCN  and N$_2$D$^+$  were  presented by  \citet{Yen2016}  using a  new
stacking technique to  enhance the signal to noise
(S/N) of the radial profile confirming the extent of  DCO$^+$. The observational study of \citet{Huang2017} found
that  both DCO$^+$  and H$^{13}$CO$^+$  show an  emission break
around 200 AU.
 
ALMA continuum emission in Band 6  of HD163296 shows a rich structure
of    rings    and    depressions    extending   up    to    230    AU
\citep{Isella2016,Zhang2016}. This  substructure could have  an impact
on   the  distribution   and   chemistry  of   certain  species   (See
sec.~\ref{sec:dis}). The gas,  traced by C$^{18}$O, extends  up to 360
AU well  beyond the  extent of the  millimeter continuum  emission and
scattered    light   has    been    detected    up   to    $\sim$500 AU
\citep{Garufi2014,Wisniewski2008}.    By
contrasting the observations  of key deuterated species  to a physical
model of HD163296 we  hope to determine  the origin  of their formation  and their
relation to the location of  the CO snowline.  We carried out observations
towards this  disk using ALMA  Band 6  and obtained spectral  cubes of
N$_2$D$^+$ ($J$=3--2), DCO$^+$ ($J$=3--2) and DCN ($J$=3--2).

The goal of this study is to constrain the radial location of these 
deuterated  species in  the disk  surrounding  the Herbig  Ae star  HD
163296 that trace different deuteration pathways.
In   Section   \ref{sec:obs}   we   present   our   data   and   their
reduction. Section \ref{sec:res} shows the spatial characterization of
the  line  emission.   Section \ref{sec:mod}  contains  our  modeling
approach  and the  derived parameters.  Section \ref{sec:dis} discusses
the  validity of  our models  and  an interpretation  of our  observed
molecules  as a  tracer for  different deuteration  pathways and  their
relation to the CO snowline.  Finally in section \ref{sec:con} we give
our conclusions.


\section{Observations}
\label{sec:obs}

We carried out observations of the disk surrounding the Herbig Ae star
HD163296    ($\rm\alpha_{2000}$     =    ${\rm    17^h56^m51{\fs}21}$,
$\delta_{2000}$ =  $-21^\circ57'22{\farcs}0$) using the  Atacama Large
(sub-)Millimeter Array (ALMA)  in Band 6 as  a part of Cycle  2 on the
27th, 28th and  29th of July 2014 (project  2013.1.01268.S). The total
integration time on source was 4 hours and 43 min with thirty-three 12
m antennas. The  correlator set up had 7  different spectral windows:
three of them contain H$_2$CO lines \citep{Carney2017} while three
others  are  centered  on  the rest  frequencies  of  DCO$^+$  ($J$=3--2),
N$_2$D$^+$ ($J$=3--2)  and DCN ($J$=3--2), all  of them at a  resolution of 61
kHz.  A final  SPW contains  wideband  (2 GHz)  continuum centered  at
232.989  GHz. The  quasar J1733-1304  was used  as gain,  bandpass and
total flux calibrator with 1.329 Jy on the lower sideband and 1.255 Jy 
in the upper sideband.

\begin{figure*}[!htb]
  \captionsetup{width=.9\textwidth}
\begin{subfigure}{0.95\textwidth}
        \centering
\includegraphics[width=.9\textwidth]{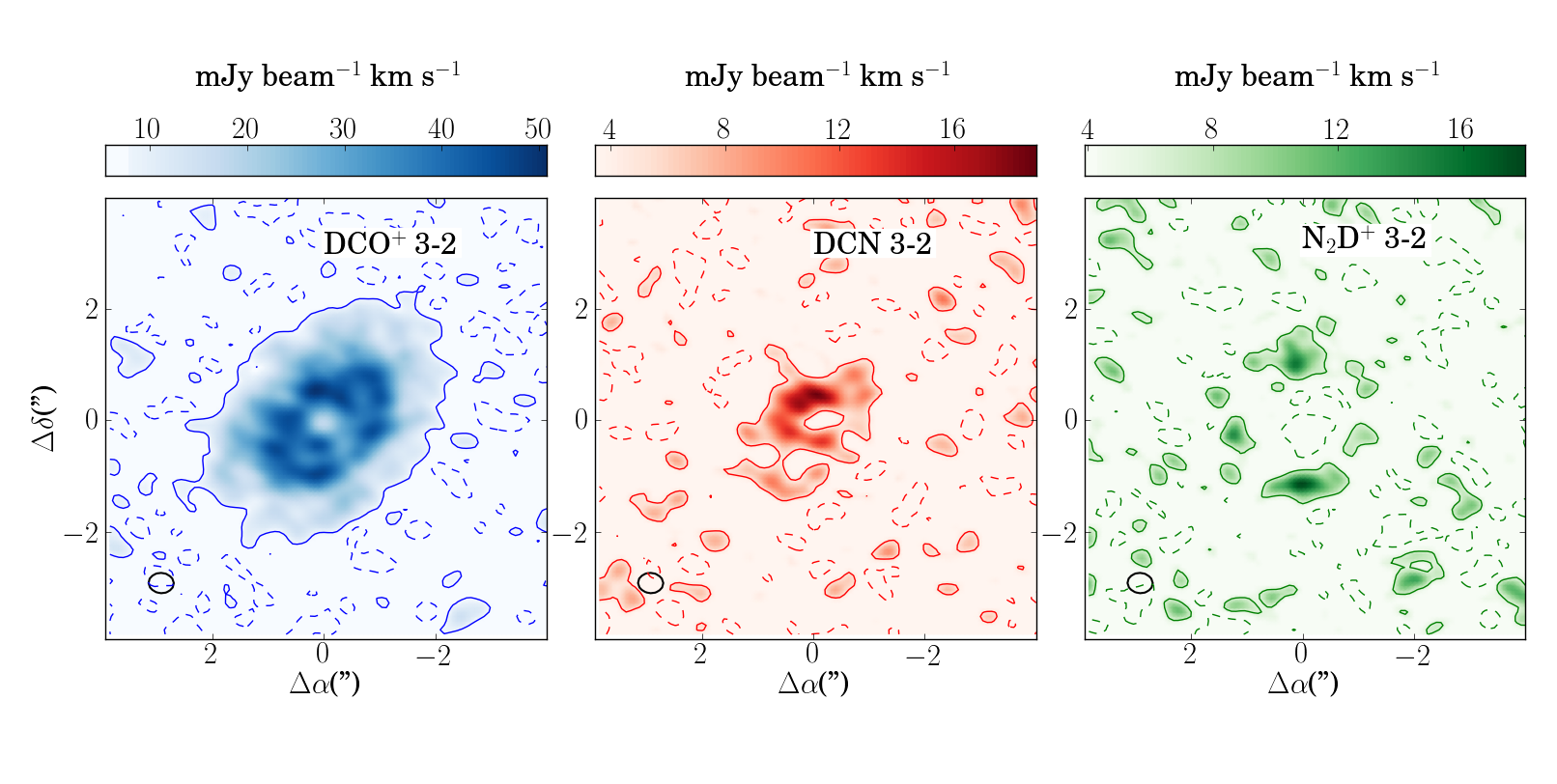}
\end{subfigure}
\begin{subfigure}{0.95\textwidth}
        \centering
\includegraphics[width=.9\textwidth]{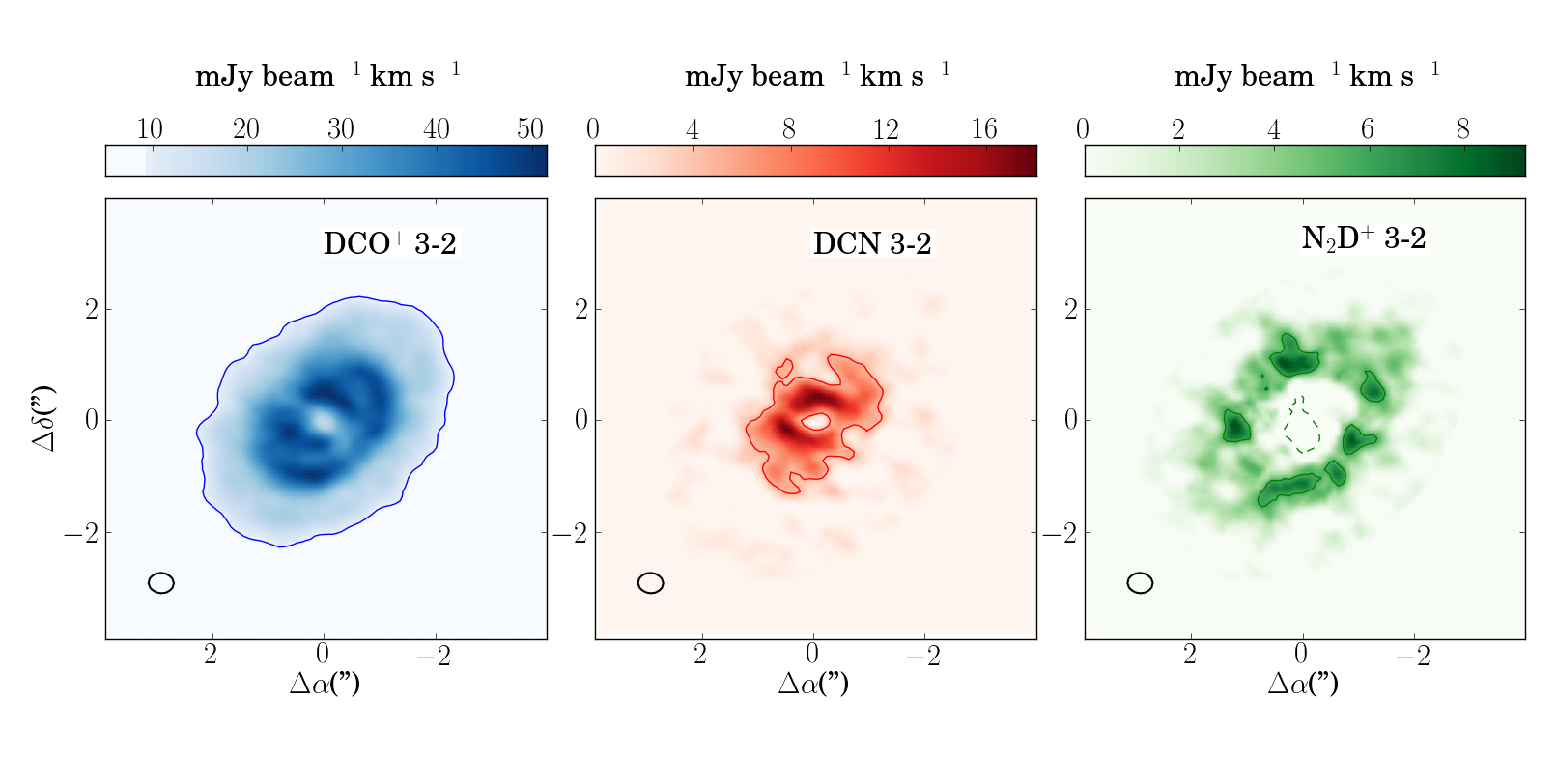}
\end{subfigure}
\caption{Integrated intensity  maps of DCO$^+$,  N$_2$D$^+$ and  DCN 
with (lower panels) and without (upper panels) a Keplerian mask as explained in Appendix~\ref{appendix}. The
resulting synthesized  beams of  0${\farcs}$53$\times$0${\farcs}$42 for
DCO$^+$ and DCN, and 0${\farcs}$50$\times$0${\farcs}$39 for N$_2$D$^+$
using  natural  weighting  are  shown  at  the  left  bottom  of  each
map. Contours  are 1-$\sigma$ levels,  where $\sigma$ is  estimated as
the rms of an emission free region in the sky. }
\label{fig:mom_all_lines}
\end{figure*}

  The data  were calibrated following  the standard CASA  reduction as
provided in the calibration scripts by ALMA.  Baselines in the antenna
array configuration correspond to a  range in $uv$-distance of 20--630 k$\lambda$, which  translates into a beam  of $\sim$ 0${\farcs}$33. Self calibration  was applied to the data as  implemented by \citet{Carney2017}.   The DCO$^+$  (J=3--2), N$_2$D$^+$  (J=3--2) and  DCN
(J=3--2) lines were continuum subtracted in the visibility plane using
a linear fit to the line-free channels  and imaged by the CLEAN task in
CASA.  Figure~\ref{fig:mom_all_lines} shows integrated intensity
maps of each line with and without a Keplerian mask. This mask is constructed by calculating the projected Keplerian velocities of the disk and matching them with the expected Doppler shifted emission in the data's spectral cube (see Appendix \ref{appendix}).  The N$_2$D$^+$ ($J$=3--2) emission is sitting on the
edge of  a strong atmospheric feature  at high $T_{\rm sys}$.  We  fit the
continuum only using the least  noisy line-free channels from one side
of the spectra (channels 350--750) but continuum subtraction is less accurate here.

\section{Results}
\label{sec:res}

\subsection{Detections}
\label{sec:res:det}

We  successfully detected  all  of our  target  emission lines.  Table
\ref{tab:lines}  shows  a  summary  of the  line  emissions  based  on
integrated intensity maps with Keplerian masking (lower panels of Fig.\ref{fig:mom_all_lines}).

\begin{figure}[]
\centering
\begin{subfigure}{0.95\columnwidth}
        \centering
        \includegraphics[width=0.95\columnwidth]{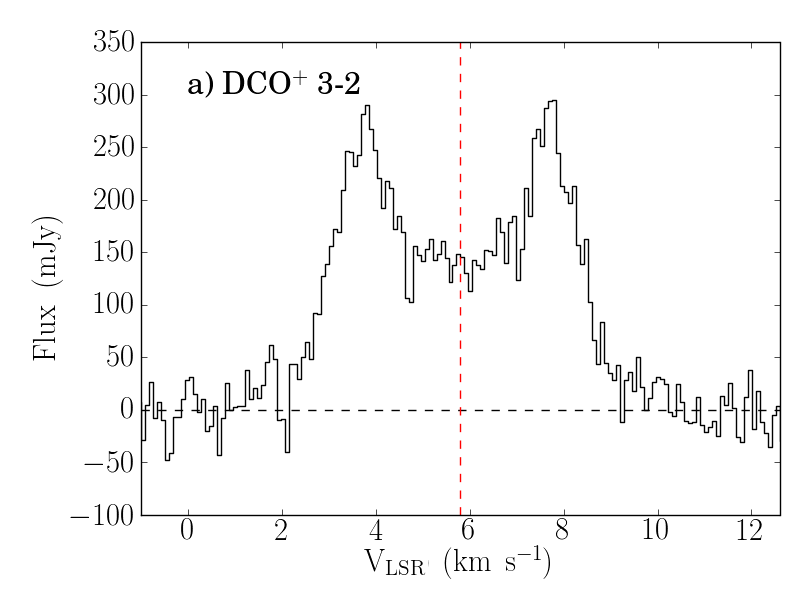}
\end{subfigure}
    \hfill
\begin{subfigure}{0.95\columnwidth}
        \centering
        \includegraphics[width=0.95\columnwidth]{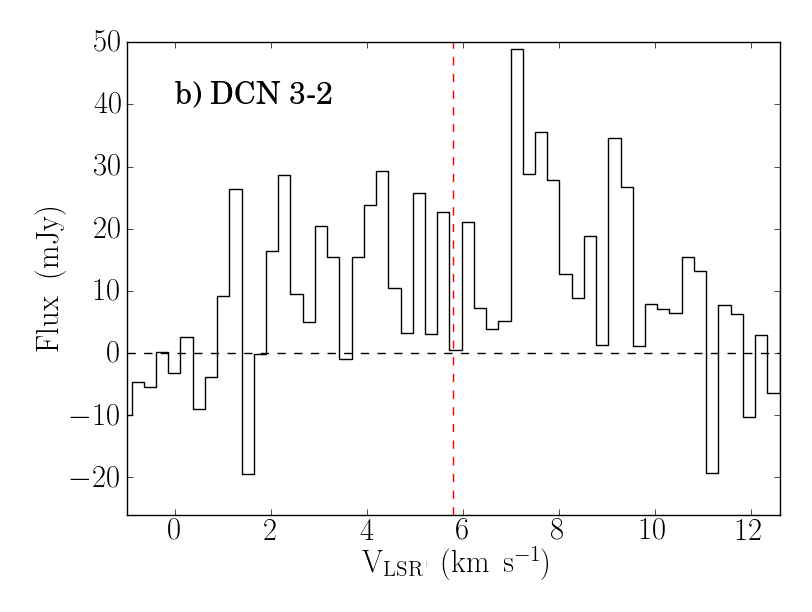}
\end{subfigure}
    \hfill
\begin{subfigure}{0.95\columnwidth}		
        \centering
        \includegraphics[width=0.95\columnwidth]{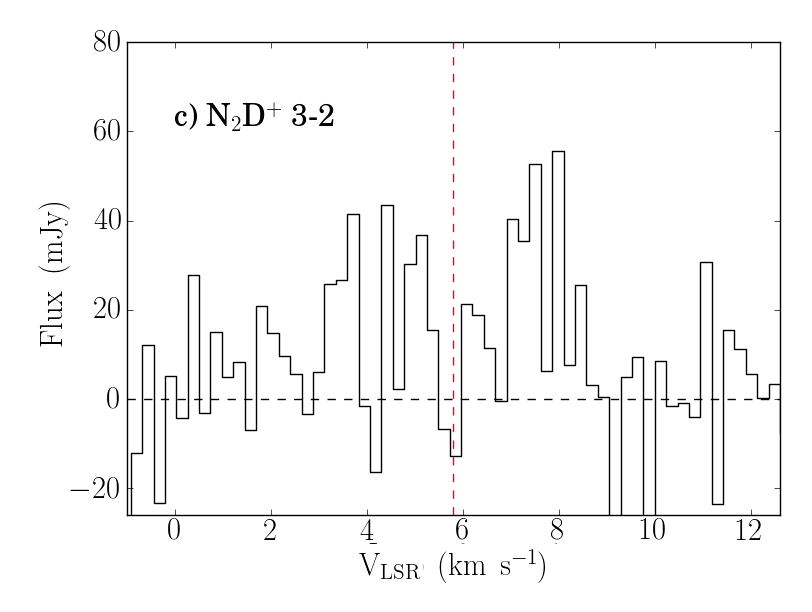}
\end{subfigure} 
	\caption{Spectra of  a) DCO$^+$ J=3--2,  b) DCN J=3--2  and c)
N$_2$D$^+$ J=  3--2 of  the disk integrated  values using  a Keplerian
mask as shown in  Fig.~\ref{fig:mom_all_lines}. The DCN and N$_2$D$^+$
spectra have been  binned to 3 times their resolution,  0.255 km/s and
0.238  km/s  respectively,  to  enhance  their  signal-to-noise  (S/N)
ratio. }
  \label{fig:Spectra} 
\end{figure}

Figure~\ref{fig:Spectra} shows  the double  peaked spectra  of
DCO$^+$ ($J$=3--2), DCN ($J$=3--2) and N$_2$D$^+$ ($J$=3--2) at peak fluxes
of  294 mJy,  48 mJy and  54  mJy (without binning); corresponding  to  detection s of
11$\sigma$, 5$\sigma$  and 4$\sigma$  respectively. These spectra correspond to an aperture in the sky equal to the 1$\sigma$ contour of DCO$^+$ in Fig.~\ref{fig:mom_all_lines} for all lines. A  simple Gaussian
fit of the DCO$^+$ line  profile gives a FWHM of $\sim$6.0  km s$^{-1}$ and +5.8 km  s$^{-1}$
offset or systemic velocity consistent with values in the literature.

\begin{table*}[]
  \caption{\label{tab:lines}Summary of our  line observations.}
\centering 
  \begin{tabular}{c  c  c   c  c}  
\hline\hline 
Line transition     & Frequency & Integrated Intensity & Beam                            & Channel rms \\ 
~                   & (GHz)     & (mJy km s$^{-1}$)    & ~                               & (mJy/beam)  \\  \hline 
DCO$^+$ J=3--2      & 216.112   & 1270.45$\pm$  5.8      & 0$\farcs$53$\times$ 0$\farcs$42 & 3.25        \\ 
DCN  J=3--2         & 217.238   & 104.4$\pm$  5.6        & 0$\farcs$53$\times$0$\farcs$42  & 3.11        \\ 
N$_2$D$^+$ J=  3--2 & 231.321   & 61.6$\pm$ 7.5        & 0$\farcs$50$\times$0$\farcs$39 & 3.37
  \end{tabular}\tablefoot{  Line  parameters  of  CLEAN  images  using
natural weighting. The velocity integrated fluxes and their respective
errors are  calculated using  a Keplerian  mask as  explained in
Appendix \ref{appendix}. }
\end{table*}

From the  integrated intensity maps, shown in Fig.~\ref{fig:mom_all_lines},  N$_2$D$^+$ seems to  be emitting
from  a broad  ring  with its  peak  flux at  9.3  mJy beam$^{-1}$  km
s$^{-1}$ and a velocity integrated line intensity of 61.6$\pm$ 7.5 mJy
km s$^{-1}$. On  the other hand, DCN emission is  more compact, arising
from within the N$_2$D$^+$ ring. As already noted by \citet{Huang2017}
in their observations, we do not  see clear evidence for an offset from
the  center  as reported  by  \citet{Yen2016}.  The emission  peak  is
somewhat    shifted   northeast    from    what    is   observed    by
\citet{Huang2017}. The total DCN velocity integrated line intensity is
104.4$\pm$ 5.6  mJy km s$^{-1}$  and the  peak flux of  its integrated
intensity map is at 17.3 mJy beam$^{-1}$ km s$^{-1}$.  DCO$^+$ extends
radially further inwards than what was observed by \citet{Mathews2013}
as confirmed by \citet{Qi2015} and \cite{Huang2017} within both of the
N$_2$D$^+$ and  N$_2$H$^+$ emission rings. The  DCO$^+$ emission peaks
at  51.5 mJy  beam$^{-1}$ km  s$^{-1}$  northwest as  noticed by  both
\citet{Yen2016} and \citet{Huang2017}, and  has a velocity integrated
line intensity of 1270.5$\pm$ 5.8 mJy km$^{-1}$.

\begin{figure}[]
  \centering
  \includegraphics[width=0.95\columnwidth]{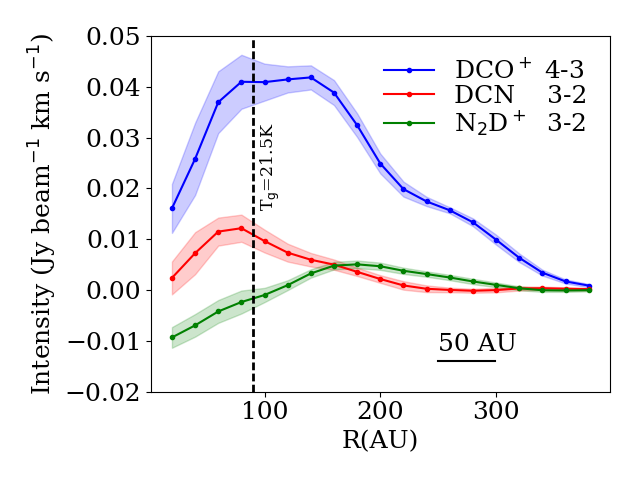}
  \caption{Radial profiles  of the  integrated
intensity  maps shown  in  Fig.~\ref{fig:mom_all_lines}. The  shadowed
color  area represents  the 3$\sigma$  errors, where  $\sigma$ is  the
standard deviation  in one elliptical  annulus. The black  dashed line
corresponds   to   the   location    of   the   CO   snowline\citep[90
AU,][]{Qi2015}.}
\label{fig:Line_profiles}
\end{figure}

Figure~\ref{fig:Line_profiles} shows  the average radial profile  of the
integrated  intensity   maps  of   Fig.~\ref{fig:mom_all_lines}.  This
profile  is  constructed  taking   the  average  value  of  concentric
ellipsoid  annuli  and  their  error  is  taken  from  the  standard
deviation.  Note that the projected linear resolution is lower along the semi-minor axis than along the semi-major axis and therefore the resulting spatial resolution of the radial profiles is poorer than the synthesized beam.  The N$_2$D$^+$ and  DCN emission peak
at $\sim$160  AU and  $\sim$60 AU  respectively. The  DCO$^+$ emission
shows 3  three peaks at ~60  AU, ~130 AU,  ~250 AU, and extends  up to
~330  AU.  Both  the  DCN  and the  DCO$^+$  radial  profiles  show  a
depression  towards  the center  of  the  disk  that is  discussed  in
Sec.~\ref{sec:dis}.

\subsection{Column densities and deuterium fractionation}

We can  get an estimate of  the disk-averaged column densities  of the
observed  species   if  we   consider  the  analytical   formula  from
\citet{Remijan2003} for optically thin emission,
\begin{equation}    \label{eq:Remijan}     N=\frac{2.04\int\Delta    I
d\nu}{\theta_a\theta_b}\frac{Q_{\rm  rot}{\rm exp}(\frac{\rm  E_u}{\rm
T_{ex}})}{v^3\langle     S_{\rm     ij}\mu^2\rangle}\times     10^{20}
cm^{-2}\rm{,}
\end{equation} where:  $\theta_a$ and $\theta_b$  (arcsec) corresponds
to  the  semi-major  and  semi-minor axis  of  the  synthesized  beam,
$\int\Delta  I  d\nu$  is  the  total line  flux  (Jy  beam$^{-1}$  km
s$^{-1}$) and T$_{ex}$  (K) is the excitation  temperature. The partition
function  (Q$_{rot}$), upper  energy  level (E$_u$  K), line  strength
(S$_{ij}$) and  dipole moment ($\mu$  Debye) were taken from  the CDMS
database.

We adopt two different disk-averaged excitation temperatures for each of the molecules: 10 and 25 K for N$_2$D$^+$ and 25 and 80 K for DCN and  DCO$^+$. N$_2$D$^+$ should be abundant at temperatures $\lesssim$ 25 K where CO is frozen-out  \citep{Qi2015} and at temperatures $\gtrsim$ 10 K where the low deuteration channel starts to be active. On the other hand, DCN and DCO$^+$ start to be abundant at higher temperatures where the high temperature pathway starts to be active (80K, see below Sec.~\ref{sec:mod:chem}). Table \ref{tab:column_density} summarizes our
column density estimates for the three deuterated species using the values of the velocity integrated line intensities from Table \ref{tab:lines}. In general,
our estimates  do not  differ by more  than a factor  of 2  within the
excitation temperature ranges.  Table~\ref{tab:ratios} shows estimated
disk-averaged  deuterium fractionation (D$_{\rm f}$) values for each of our species  using  the disk-averaged  column
densities from Table~\ref{tab:column_density}.  We take the velocity integrated
flux values of past line  detections of H$^{13}$CO$^+$ $J$=3--2 (620 mJy
beam$^{-1}$ km  s$^{-1}$), H$^{13}$CN  $J$=3--2 (170 mJy  beam$^{-1}$ km
s$^{-1}$) \citep{Huang2017},    N$_2$H$^+$ $J$=3--2 (520 mJy  beam$^{-1}$ km
s$^{-1}$)   \citep{Qi2015}    and
Eq.~\ref{eq:Remijan} to derive  disk-average column densities assuming
a $^{12}$C/$^{13}$C ratio of  69 \citep{Wilson1999}. The $^{12}$C/$^{13}$C ratio can vary in disks between different C-bearing species \citep{Woods2009}.  Higher or lower $^{12}$C/$^{13}$C ratios for these species change D$_{\rm f}$ linearly. These D$_{\rm f}$
values are  only lower limits  as the  species might not  be spatially
co-located.

\begin{table}
 \caption{\label{tab:column_density}Column
density estimates for  different excitation temperatures.}  \centering
  \begin{tabular}{c c c c } 
\hline\hline  
T$_{ex}$ & N$_{\rm DCO^+}$     & N$_{\rm DCN}$       & N$_{\rm N_2D^+}$     \\
~                      & (cm$^{-2}$)         & (cm$^{-2}$)         & (cm$^{-2}$)          \\  \hline
 10 K         & -                   & -                   & 2.5 $\pm 0.3\times 10^{11}$ \\
 25 K         & 1.68$\pm 0.01\times 10^{12}$ & 2.9$\pm 0.2\times 10^{11}$ & 1.6 $\pm 0.2\times 10^{11}$ \\ 
 80 K	       & 2.56$\pm 0.01\times 10^{12}$ & 5.2$\pm 0.3\times 10^{11}$ & -
  \end{tabular}
\end{table}
\begin{table}
  \caption{\label{tab:ratios}Deuterium fractionation
estimates for  different excitation temperatures.}  \centering  
  \begin{tabular}{c c c c } 
\hline\hline   
T$_{ex}$ & ${\rm DCO^+/HCO^+}$ &  ${\rm DCN/HCN}$  &  ${\rm N_2D^+/N_2H^+}$                                \\
 \hline
 10 K   & -   & -  & 0.34$\pm$0.15 \\
 25 K   & 0.05$\pm$0.01 & 0.02$\pm$0.01  & 0.45$\pm$0.21\\ 
 80 K   & 0.06$\pm$0.01 & 0.02$\pm$0.01  & --\\
  \end{tabular}
\end{table}

\section{ Parametric modeling}
\label{sec:mod}
\subsection{Deuterium chemistry}
\label{sec:mod:chem}

 The D/H  ratio is  specially
useful to  constrain the  physical conditions of  protoplanetary disks
since an  enhancement in  this ratio  is a  direct consequence  of the
energy barrier  of the reactions  that deuterate simple  molecules. We
can distinguish three different key reactions that introduce deuterium
into    the   most    abundant   species    in   protoplanetary    disks
\citep{Gerner2015,Turner2001},
\begin{subequations}
\begin{equation} \rm H_3^+ +HD\leftrightharpoons H_2D^++H_2+230K{\rm,}
\label{eq:low_T}
\end{equation}
\begin{equation} \rm CH_3^+ +HD\leftrightharpoons CH_2D^+ +H_2+370K{\rm,}
\label{eq:high_T}
\end{equation}
\begin{equation} \rm C_2H_2^+ +HD\leftrightharpoons C_2HD^+ +H_2+550K{\rm.}
\label{eq:high_T_2}
\end{equation}
\end{subequations}

The right-to-left  reaction of Equation~\ref{eq:low_T}  is endothermic
and  strongly enhances  the D/H  ratio of  H$_2$D$^+$, and  species that
derive   from    it,   in   temperatures   ranging    from   10-30   K
\citep{Millar1989,Albertsson2013}.   This  regime corresponds  to  the
so-called  low  temperature  deuteration channel.   In  contrast,  the
right-to-left reactions of  Eq.~\ref{eq:high_T} and \ref{eq:high_T_2},
involving  light  hydrocarbons,   effectively  enhance  the  deuterium
fractionation in warmer temperatures ranging from 10-80 K . This regime
corresponds to the high temperature deuteration channel.

N$_2$D$^+$  forms  mainly  through  the  low  temperature  deuteration
channel via ion-molecule reaction \citep{Dalgarno1984},
\begin{equation} \rm H_2D^++N_2\longrightarrow N_2D^++H_2.
\label{eq:N2Dp}
\end{equation}
Hence the  formation of N$_2$D$^+$ is  expected to be enhanced  at the
same temperature  range as  H$_2$D$^+$.

DCN is  formed out of  two main reactions  involving the low  and high
temperature   channels  with   the   latter   being  dominant   (66\%)
\citep{Turner2001,Albertsson2013}.   Early  works   identify  the  low
temperature  channel  as  the   key  gas-phase  formation  of  DCO$^+$
\citep{Watson1976,Wootten1987}.     This     pathway    starts    from
Eq.~\ref{eq:low_T} followed by the reaction
\begin{equation} \rm H_2D^++CO\longrightarrow DCO^++H_2\rm{.}
\label{eq:DCOp_1}
\end{equation} However, recently modeling efforts by \citet{Favre2015}
show  that the  high temperature  channel may  be an  active formation
pathway.  If we  consider Eq.~\ref{eq:high_T},  DCO$^+$ can  be formed
via
\begin{equation} \rm CH_2D^++O\longrightarrow DCO^++H_2\rm{.}
\label{eq:DCOp_2}
\end{equation}

\subsection{Motivation}
\label{sec:mod:mot}
\begin{figure}[]
  	\centering
   	\includegraphics[width=0.95\columnwidth]{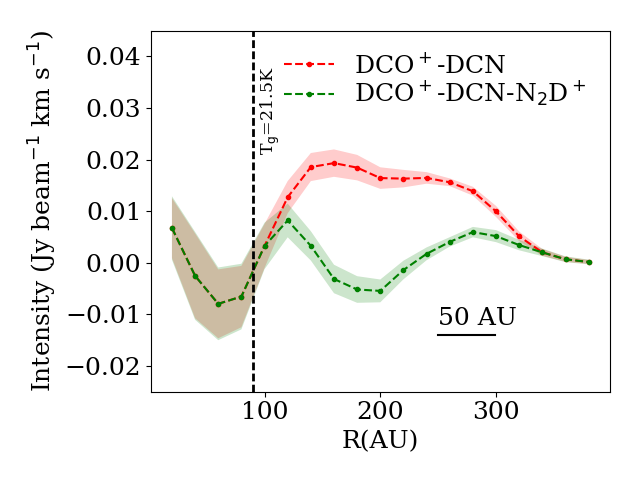}
	\caption{Residuals of the DCO$^+$ radial
profile  (as  shown  in Fig.~\ref{fig:Line_profiles})  by  subtracting
first the  DCN radial  profile and then the N$_2$D$^+$
radial profile. The shadowed  color area represents 1-$\sigma$ errors,
where   $\sigma$  is   the  standard   deviation  in   one  elliptical
annulus. The black  dashed line corresponds to the location  of the CO
snowline (90 AU).}
  	\label{fig:Line_profiles_subtracted} 
\end{figure}

Past observations of DCN in protoplanetary disks show centrally peaked distributions \citep{Qi2008,Huang2017}. This supports the idea of DCN being mainly formed through the high temperature deuteration pathway. If N$_2$D$^+$ and DCN are  tracing the  low and high temperature
deuteration pathways, respectively, we  can think of the DCO$^+$ emission  as a linear
combination                 of                 DCN                 and
N$_2$D$^+$. Figure~\ref{fig:Line_profiles_subtracted}  shows, as an illustration, the DCO$^+$
radial profile subtracted first by the DCN emission scaled by a factor
of 3.7 and then  by the N$_2$D$^+$ emission scaled by  a factor of 4.9. These factors  can be interpreted as the  ratio of their
abundances.   The  first  ring  at  ~115 AU  in  the  residuals,  after
subtracting  both  the  DCN  and  N$_2$D$^+$  radial  profile,  can  be
interpreted  as DCO$^+$  that is  formed inside  the CO  snowline.  If
N$_2$D$^+$  is formed  exclusively outside  the snowline,  its emission
peak  will be  slightly shifted  outwards \citep{vantHoff2016,Zhang2017} and  its
subtraction will reveal DCO$^+$ formed  by the low temperature channel
in the regions where CO is still present in the gas phase . The second
residual ring  at ~280 AU  indicates a  third regime where  DCO$^+$ is
present in  the gas phase that  does not correlate with  the deuteration
pathways regimes probed by DCN and N$_2$D$^+$.

We intend to characterize the  three different regimes described above
using a reasonable  physical model for the disk and  simple models for
the  column  density of  our  species.  In  the following  section  we
describe the adopted physical model  and the parametric abundances for
N$_2$D$^+$, DCN and DCO$^+$ that are used to fit the data.
\subsection{Physical model}
\label{sec:mod:phy}

 We adopt the  physical model used by  \citet{Mathews2013}. This model
was  constructed by  approximating  the  semi-parametric modeling  of
\citet{Qi2011} fitting the SED and the extent of their mm observations
.The density structure is defined by

\begin{equation*}
\Sigma_d(R)=
\begin{cases}
\Sigma_C~\left(\frac{R}{R_c}\right)^{-1}{\rm
  exp}\left[-\left(\frac{R}{R_c}\right)\right] & \text{if }R \geq R_{\rm rim} \\
0 & \text{if }R < R_{\rm rim ,}
\end{cases}
\end{equation*} 
where $\Sigma_{\rm  C}$ is determined  by the total disk  mass $M_{\rm
disk}$ (0.089 M$_{\odot}$  using a gas-to-dust ratio of 0.0065), $R_{\rm C}$
(150 AU) is the characteristic radius and $R_{\rm rim}$ (0.6 AU) is the
inner rim of the disk. The vertical structure is treated as a Gaussian
distribution with an angular scale height defined by
\begin{equation*}
h(R)=h_{\rm C} \left(\frac{R}{R_{\rm C}}\right)^\psi ,
\end{equation*} 
where $\psi$ (0.066) is the flaring  power of the disk and $h_{\rm C}$
is  the angular  scale  height at  the  characteristic radius  $R_{\rm
C}$. The parameter $h_{\rm C}$ takes different two values for the dust
vertical   distribution   and   two   more  for   the   gas   vertical
distribution. The parameters $h_{\rm   small}$(0.08)  and  $h_{\rm   large}$  (0.06)
describe   the  distribution   of   small  and   large  dust   grains,
respectively, while $h_{\rm  main}$ (0.1) and $h_{\rm  tail}$ (0.2) 
describe the  main bulk distribution  of gas  in the mid-plane  of the
disk and the tail  of gas that continues the upper  regions of the disk
surface  \citep[See appendix  A of][]{Mathews2013}.  The gas  temperature
profile  was  computed  by  the   2D  radiative  transfer  code  RADMC
\citep{Dullemond2004b} assuming T$_{\rm gas}=$T$_{\rm dust}$.  This code  receives as  an input  the stellar
parameters  as  listed in  \citet{Mathews2013}  and  the dust  density
distribution.

\subsection{N$_2$D$^+$, DCN and DCO$^+$ abundance models}
\label{sec:mod:abu}  

DCN is formed mainly (66\%) through deuterated light hydrocarbons such
as  CH$_2$D  \citep[see  Fig.  5c  of][]{Turner2001}.   This  reaction
starts  with   the  deuteration   of  CH$_3^+$,  the   so-called  high
temperature deuteration  pathway. For  an enhancement of  CH$_2$D$^+$ over
CH$_3^+$  temperatures  should not  exceed  $\sim$10--80 K.   We use  a
simple  toy  model  for  DCN  taking the  same  approach  as  for  the
N$_2$D$^+$  model  with  three  free-parameters: an  an  inner  radius
$R_{\rm in}$, beyond it is sufficiently cold for an enhancement of the
CH$_2$D/CH$_3^+$ ratio, an  outer radius $R_{\rm out}$,  and a constant
abundance $X_{\rm high}$.

N$_2$D$^+$ is  formed by  the reaction of  N$_2$ with  H$_2$D$^+$.  We
expect  considerable  abundances of  N$_2$D$^+$  only  outside the  CO
snowline because its parent molecule, H$_3^+$, is readily destroyed by
proton exchange with CO. This is also true for its non-deuterated form
N$_2$H$^+$.   We use  a simple  ring  model to  constrain the  spatial
distribution  of   gas  N$_2$D$^+$.   The  model   consists  of  three
free-parameters: an inner  radius $R_{\rm in}$ beyond  the CO snowline
and  where conditions are sufficiently  cold   for  a   substantial  enhancement   of  the
H$_2$D$^+$/H$_3^+$ ratio, an outer radius  $R_{\rm out}$, and a constant
abundance $X_{\rm low}$.

We model the  distribution of DCO$^+$ as  three separate contributions
from both  deuteration channels and a  third region in the  outer disk
motivated  by  the radial  profile  of  its integrated  intensity  map
(Fig.~\ref{fig:Line_profiles_subtracted}).   Our model  uses a  set of
seven parameters  describing three  regions: an inner radius ($R_{\rm
in}$),  two radial breaks  ($R_1$, $R_2$)  and  three constant  abundances
($X_{\rm high}$, $X_{\rm low}$, $X_{3}$) for  the inner, the middle and the outer emission region. Figure~\ref{fig:toy_models} shows abundances and column density profiles of these simple models.

\subsection{Line excitation}

Instead  of a  full radiative  transfer modeling  we opted  for a  more
simplistic  approach  by  considering estimates  of  a  characteristic
excitation temperature  as a function  of the distance to  the central
star  to calculate  the  resulting line  emission  given an  abundance
profile. We used  LIME \citep[v1.5]{Brinch2010}, a 3D radiative transfer  code in non-LTE
that can produce line and continuum radiation from a physical model to
estimate a characteristic excitation  temperature as a function of radius throughout the disk for the  observed  transitions:
DCO$^+$ 3--2, DCN 3--2 and N$_2$D$^+$  3--2. We use the physical model
described above  and a  constant abundance  equal to  the disk-averaged
found in Sec.~\ref{sec:res} assuming $T_{\rm ex}=25$ K.  LIME outputs population
levels in a grid of 50000 points which  are randomly distributed in $R$ using a
logarithmic scale.   Establishing a convergence  criteria encompassing
all of  the grid points  is difficult. We  manually set the  number of
iterations  to 12  and  confirm convergence  by comparing  consecutive
iterations.  We use  the rate coefficients from the  Leiden Atomic and
Molecular                                                     Database
\citep{LAMBDA_database2005}\footnote{www.strw.leidenuniv.nl/$\sim$moldata/}. For
N$_2$D$^+$  and DCN  we use, respectively,  the  N$_2$H$^+$ rate coefficients, which  are adopted  from
HCO$^+$    \citep{Flower1999},     and  the  HCN     rate    coefficients
\citep{Dumouchel2010} since the transitions between the non-deuterated
and deuterated forms  do not differ significantly. For  DCO$^+$ we use
the same  collision rates as the  ones for HCO$^+$ and  the Einstein A
coefficients taken from  CDMS and JPL.

To construct the radial excitation temperature profile we take  the average excitation
temperature of  the points below  $z<10\times h(R)$, where  $h(R)$ is
the scale height of our adopted  physical model. These temperature profiles are
shown in red in Fig.\ref{fig:toy_models}.  A drawback of this approach
is  that assuming  an  isothermal  vertical structure  of  the  temperature
profile  might not properly  describe the vertical  region over which
the molecules extend.  

\subsection{Abundance estimates}
\begin{figure*}[]
\centering
\begin{subfigure}{0.49\textwidth}
        \centering
        \includegraphics[width=0.98\textwidth]{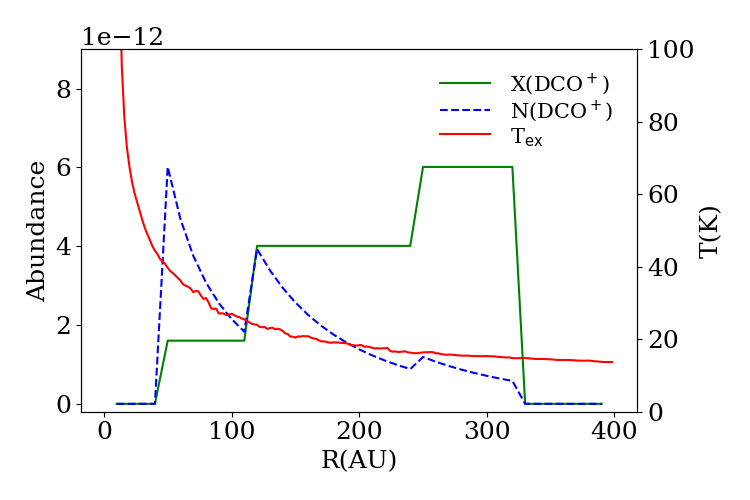}
\end{subfigure}
\begin{subfigure}{0.49\textwidth}
        \centering
        \includegraphics[width=0.98\textwidth]{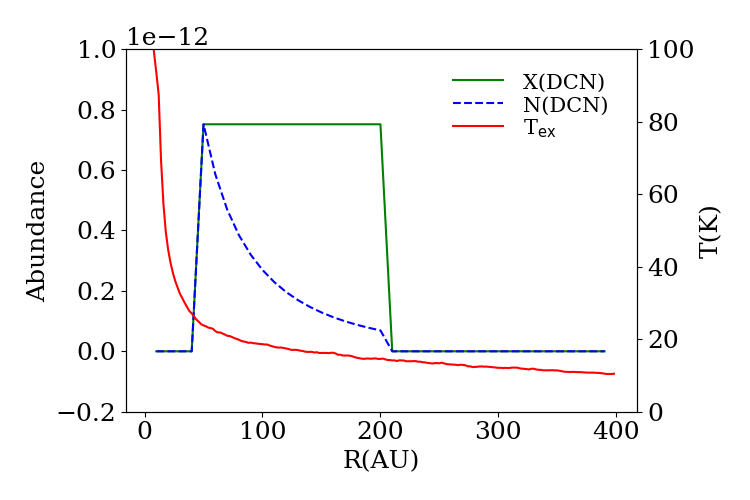}
\end{subfigure}
\begin{subfigure}{0.49\textwidth}		
        \centering
        \includegraphics[width=0.98\textwidth]{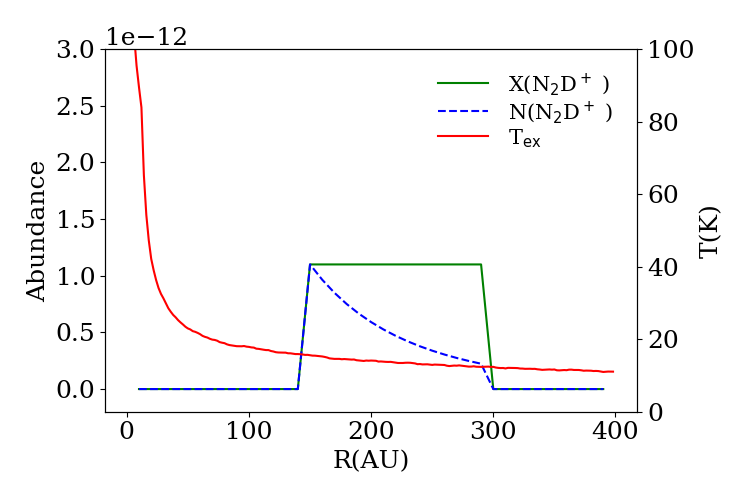}        
\end{subfigure} 
\caption{Schematics of the abundance models for the  transitions of DCO$^+$  $J$=3--2, DCN
$J$=3--2 and N$_2$D$^+$  $J$=3--2. The dashed blue line  corresponds to an
scaled profile  of the column density of the abundance model for the purpose of illustration.  The red
continuous  line corresponds  to  the  assumed excitation  temperature
profile.}
\label{fig:toy_models}
\end{figure*}

We compute  a radial emission profile using the
radial excitation  temperature profile  and the column  densities from
the  parametric constant  abundance  profiles, shown in green in Fig.\ref{fig:toy_models}, by  solving
Eq.\ref{eq:Remijan}.  

We then  create a 2D sky image  from the modeled
emission profile and  convolve it with the synthesized beam of the observation to
fit      the     integrated      intensity      maps     shown      in
Fig.~\ref{fig:mom_all_lines}.  Since we only fit radial column density
profiles  the  vertical  distribution   of  these  species  cannot  be
constrained by this approach. However,  past modeling of DCO$^+$ shows
that  its  vertical distribution  has  limited  effect on  constraining
radial boundaries \citep{Qi2015}.

\subsection{Best-fit parameters}	

\begin{figure*}[]
\centering
\begin{subfigure}{0.49\textwidth}
        \centering
        \includegraphics[width=0.98\textwidth]{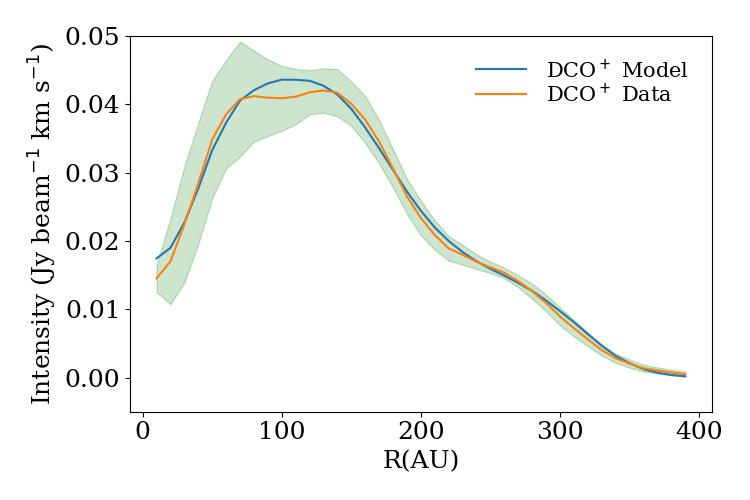}
\end{subfigure}
\begin{subfigure}{0.49\textwidth}
        \centering
        \includegraphics[width=0.98\textwidth]{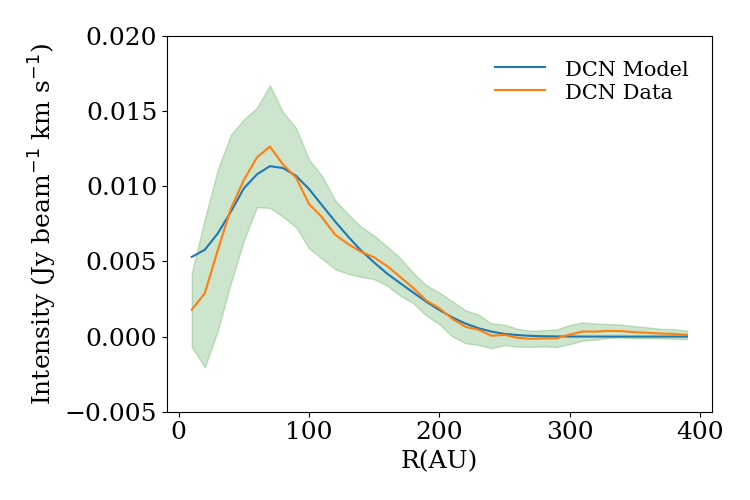}
\end{subfigure}
\begin{subfigure}{0.49\textwidth}		
        \centering
        \includegraphics[width=0.98\textwidth]{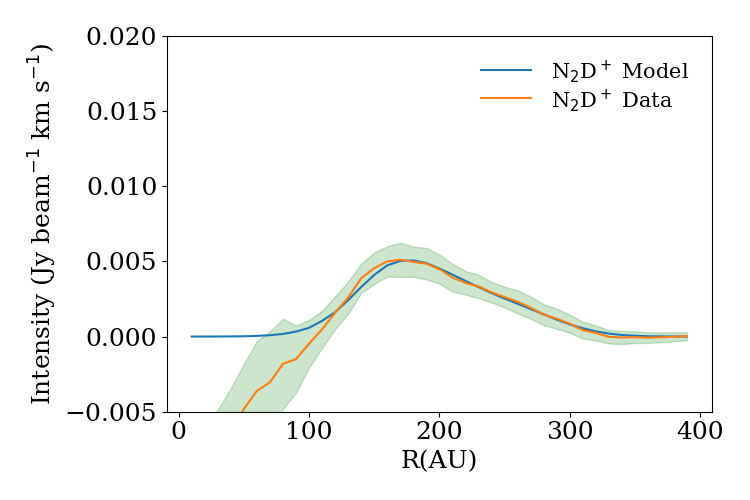}        
\end{subfigure} 
\caption{Best-fit models  for the  transitions of DCO$^+$  $J$=3--2, DCN
$J$=3--2 and N$_2$D$^+$  $J$=3--2. The error bars, shown as a filled region in green, correspond  to the standard deviation of the
values in one annulus as seen in Fig.~\ref{fig:Line_profiles}.}
\label{fig:chanel_lines}
\end{figure*}

  We                           minimize                          ${
\chi^2=(F_{\rm obs}(x,y)-F_{\rm model}(x,y))^2/\sigma(x,y)^2}$  values,  where
${F_{\rm obs}(x,y)}$ corresponds to the  data points of the integrated
intensity  maps  (as  shown  in  Fig.~\ref{fig:mom_all_lines}),  ${F_{\rm model}(x,y)}$ corresponds  to the points of  the modeled integrated
intensity map  convolved with  the synthesized beam  and $\sigma(x,y)$
corresponds to  the standard  deviation of the pixels of a concentric  
ellipsoid (as the ones used to make the radial profiles seen in 
Fig.\ref{fig:Line_profiles}) that includes the pixel $(x,y)$ . We  report best-fit parameters that correspond to
a minimum of  the explored parameter space. In the  case of DCO$^+$ we
explore different configurations of rings,  allowing them to have one,
two  or  no gaps  empty  of  material  between the  ring-like  regions.
Table~\ref{tab:summary} shows  a summary  of our preferred  models and
Fig.~\ref{fig:chanel_lines}  show  radial  profiles  of  the  best-fit
model's  column density,  integrated  intensity, estimated  excitation
temperature and the  integrated intensity of the data for  each of our
detections.

In the case of DCO$^+$ the  location of the different radial zones are
degenerate.  A  model  with  an additional  two  degrees  of  freedom,
describing  three radial rings separated by gaps,  can also reproduce the  data.  The modest spatial
resolution  of our  data  cannot distinguish  between these models.   Detailed chemical  modeling  plus  a 3D  radiative
treatment of DCO$^+$  is necessary to break the degeneracy  of our toy
model.  The  best-fit  abundances  for the  simplest  model
without any  gaps as  described in Sec.~\ref{sec:mod:abu}  are $X_{\rm
high}$=1.6$^{+0.4}_{-0.5}\times$10$^{-12}$,                    $X_{\rm
low}$=4.0$^{+1.0}_{-1.3}\times$10$^{-12}$,
$X_3$=6.0$^{+1.4}_{-2.0}\times$10$^{-12}$  and  their best-fit  radial
boundaries are $R_{\rm in}$=50$^{+5}_{-3}$ AU, $R_1$=118$^{+4}_{-5}$ AU ,
$R_2$=245$^{+3}_{-13}$ AU and $R_{\rm out}$=316$^{+3}_{-10}$ AU. We do not consider a vertical
distribution of the DCO$^+$ and DCN molecules. This results in  vertically averaged abundances which are lower limits to the maximum values expected in a full 2D treatment because in reality DCO$^+$ and DCN are absent where the gas temperature is higher than the activation temperature for their deuteration. In addition, DCO$^+$ is absent in the midplane where CO starts to freeze-out.

The best-fit  model for the  DCN emission  profile consists of  a ring
with              constant              abundance              $X_{\rm
high}$=7.5$^{+0.9}_{-0.9}\times$10$^{-13}$       between       $R_{\rm
in}$=51$^{+6}_{-6}$ AU  and $R_{\rm out}$=201$^{+15}_{-24}$  AU, where
$R_{\rm in}$ can be thought as the radius where the gas temperature is
high enough  for the reaction  described in Eq.~\ref{eq:high_T}  to be
exothermic.  The  best-fit model  for the N$_2$D$^+$  emission profile
consists    of    a    ring   with    constant    abundance    $X_{\rm
low}$=1.1$^{+0.1}_{-0.1}\times$10$^{-12}$        between       $R_{\rm
in}$=139$^{+5}_{-4}$ AU and  $R_{\rm out}$=287$^{+15}_{-21}$ AU, where
$R_{\rm  in}$  could  be  tracing  the  CO  midplane  snowline  radial
location.

\begin{table}
  \caption{\label{tab:summary}Best-fit model parameters}  \centering 
  \begin{tabular}{l c c c } 
\hline\hline  
Parameters       & DCN $J$=3--2            			   & N$_2$D$^+$ $J$= 3--2    & DCO$^+$ $J$=3--2        \\ \hline 
$R_{\rm in}$     & 51$^{+6}_{-6}$ AU     			   & 139$^{+5}_{-4}$ AU    & 50$^{+5}_{-3}$ AU     \\
$R_1$             & ~                    			   & ~                     & 118$^{+4}_{-5}$ AU    \\
$R_2$             & ~                     			   & ~                     & 245$^{+3}_{-13}$ AU   \\
$R_{\rm out}$    & 201$^{+15}_{-24}$ AU  			   & 287$^{+15}_{-21}$ AU  & 316$^{+3}_{-10}$ AU   \\
$X_{\rm high}$   & 7.5$^{+0.9}_{-0.9}\times$10$^{-13}$ & ~                     & 1.6$^{+0.4}_{-0.5}\times$10$^{-12}$ \\
$X_{\rm low}$    & ~                     			   & 1.1$^{+0.1}_{-0.1}\times$10$^{-12}$ & 4.0$^{+1.0}_{-1.3}\times$10$^{-12}$ \\
$X_3$            & ~                     			   & ~                     & 6.0$^{+1.4}_{-2.0}\times$10$^{-12}$ \\
  \end{tabular}\tablefoot{ Best-fit  parameters of models as  shown in
Fig.~\ref{fig:chanel_lines}.  DCO$^+$ values  correspond to  the model
with two  gaps. The  formal errors are calculated at  90\% confidence
levels. }
\end{table}

\section{Discussion}
\label{sec:dis}

\subsection{The inner depression}

Our best-fit models find an inner drop in emission at $\sim$50 AU for both the
DCN  and  DCO$^+$  line  emission.  This  could  be  tracing  the  gas
temperature required  for the high-temperature deuteration  channel to
be active  (10-80 K) or an  optically thick continuum region  at small
radii. Recent  observations of  C$^{18}$O, $^{13}$CO and  continuum at
1.3 mm \citep{Isella2016}  with ALMA in higher  spatial resolution also
show a central depression in C$^{18}$O and $^{13}$CO radial intensity profiles. These CO
isotopes  are  optically  thin  and   trace  the  gas near  the
midplane. If the dust  becomes optically thick it could hide
the  line emission  coming from  deeper layers  towards the  midplane.
Subtracting the  continuum estimate in  these regions then  creates an
apparent absence of emission. An  alternative explanation  is  inherent to  the  chemistry of  both
species. DCN is  readily destroyed by simple ions such  as H$_3^+$ and
H$^+$ \citep{Albertsson2013}.   DCO$^+$ can be destroyed  if H$_2$O is
desorbed from the  grains at inner radii. It is  unlikely that both of
these effects  correlate at  the same  radii (R$_{\rm  in}\sim$50 AU).

The peak value of  the continuum at the center of the  disk is 167 mJy
beam$^{-1}$ corresponding to  a brightness temperature of 25  K at 1.3
mm. This  is inconsistent with  the expected dust  temperature derived
from  SED  models  for  optically  thick  dust. The disk surrounding HD163296 is known to have unresolved
ring-like structure at 1.3 mm \citep{Zhang2016,Isella2016}. The
corresponding brightness  temperature for the peak  intensity in these
observations at higher  spatial resolution is 55 K consistent with optically
thick dust.  The best-fit continuum models of \citet{Isella2016} show optically thick dust
just inside 50 AU. We conclude that our $R_{\rm  in}$ parameter is tracing this optically thick dust region.

\subsection{Limitation of DCO$^+$ as a CO snowline tracer}

N$_2$H$^+$ has  been used to  trace the CO snowline  in protoplanetary
disks  before \citep{Qi2015,Qi2013}.   We  expect  N$_2$D$^+$ to  also
trace it using the same reasoning as for N$_2$H$^+$ \citep{Huang2015}.
Our best-fit  N$_2$D$^+$ model gives a  $R_{\rm in}$ value of  141 AU,
that is 50 AU further away  than previous estimates of the CO snowline
in HD 163296. This could be  due to the apparent central depression in
the N$_2$D$^+$ radial profile. As explained in Sec.~\ref{sec:obs} this
line is  sitting on the  edge of  an atmospheric feature  resulting in
high noise and a very weak detection which makes it difficult to do a good continuum
estimation.  The central negative bowl has the same order of magnitude
as other  scattered  negative  noise  in the  image  plane  of  the
resulting integrated intensity map, but, as it is located at the center,
an azimuthal average does not represent  as good an estimate of the radial
profile in the inner disk as it does in the outer disk. We cannot rely
on our best-fit value for $R_{\rm in}$  as a probe for the CO snowline
in  the midplane,  although  a deeper  integration  of the  N$_2$D$^+$
$J$=3--2 line can still be used to constrain it.

On  the  other hand,  our  best-fit  DCO$^+$  model  value for  $R_1$  of
118$^{+4}_{-5}$ AU  agrees well  with the previous  estimate of $\sim$90 AU
given  our   resolution.  Two  effects  are   boosting  the  deuterium
fractionation of  H$_2$D$^+$ and, in consequence,  the low temperature
deuteration channel. First, the gas  temperature is low enough for the
reaction  in  Eq.~\ref{eq:low_T}  to be  mainly  exothermic.   Second,
H$_2$D$^+$ is readily destroyed by CO in the gas-phase which is frozen
out.    But   CO    must   be    present   to    form   DCO$^+$    via
Eq.~\ref{eq:DCOp_2}.  This  could  be  achieved in  a  scenario  where
DCO$^+$ is found in a thin  layer around the CO freeze-out temperature
as proposed by \citet{Mathews2013}. Without previous knowledge, DCO$^+$ 
alone cannot trace the location of the CO snowline.

\subsection{The origin of the third ring}

Motivated by the  reasoning in Section~\ref{sec:mod:mot}  we find a
best-fit  value of  245$^{+3}_{-13}$  AU for  the $R_2$  parameter in  our
DCO$^+$ model. At these radii the bulk of CO in the gas-phase is locked
up on the grains as ice.  The  main production route of DCO$^+$ in the
low-temperature deuteration route  (Eq.\ref{eq:DCOp_1}) requires CO in
the gas-phase.  This raises the question how DCO$^+$ emission arises at such large radii.

 Photodesorption of CO by non-thermal processes could explain an outer
ring  of  DCO$^+$  emission  as  discussed  in  the  case  of  IM  Lup
\citep{Oberg2015}.  However  the efficiency of this  mechanism is very
model-dependent.   Full  chemical models  of  DCO$^+$  towards DM  Tau
\citep{Teague2015}  show that,  although  there is  an enhancement  of
DCO$^+$ at  large radii  produced by the interstellar  UV field  and X-ray
luminosity, the  effect is  not as  pronounced as the  one seen  in IM
Lup. The strongly emitting outer disk seen towards IM Lup coincides with
the  extent of  the mm-size  grains. If  the smaller  grains are  also
following the steepening of the mm-size grains' surface density, UV penetration will be
enhanced leading to  an enhancement on the  DCO$^+$ emission. High-resolution   continuum   data   observed   with   ALMA   band   6
\citep{Isella2016} revealed three concentric sets of rings and gaps as
predicted by  \citet{Zhang2016}. The first depression at D$_1\sim$
53 AU and the second brightness peak  centered at B$_2\sim$
120 AU  correlate well with $R_{\rm in}$=50$^{+5}_{-3}$ AU and  $R_1$=118$^{+4}_{-5}$ AU from  our best-fit
DCO$^+$ model.  The  extent of  the continuum to about 230  AU also
correlates with  $R_2$ at 245$^{+3}_{-13}$  AU from our  best-fit DCO$^+$
model . Although  no direct consequence of the location  of the second
gap can be  drawn towards an enhancement in UV  penetration, the scales
at which  these rings are  seen in the (sub)mm-continuum  ($\sim$ 60-50
AU) are  a rough estimate of  the spatial scales where  rings and gaps
are seen  in shorter  wavelengths. UV radiation  or cosmic-ray-induced
photons  may  also  photo-dissociate  species  on  the  grains  surface
\citep{Cuppen2007} such as HDCO leading to the release of DCO$^+$. Full grain-surface chemical modelling is needed to identify the chemical reactions involved in this mechanism.

\citet{Cleeves2016} have proposed  a thermal inversion in the midplane  as a mechanism to release CO into the gas-phase at large radii .  Recently,  dust  evolution  modeling efforts  by \citet{Facchini2017}  coupled with  chemistry models on  HD163296 show
signals of  thermal inversion as  a direct consequence of  radial drift,
grain  growth and  settling. As  a  result, thermal  CO desorption  is
enhanced at  large radii where  the dust becomes warmer.   The thermal
inversion in the models of \citet{Facchini2017} occurs  at about 400 AU,  farther away than our  $R_2$ best-fit
parameter of 245$^{+3}_{-13}$ AU. However,  the radial location of the
thermal inversion is very temperature sensitive, and a slightly colder
disk could  shift its location inwards.  It is important to  note that
this effect is  only seen in models with a  low turbulence $\alpha$ of
10$^{-3}$-10$^{-4}$  and when  the  CO  ice is  mixed  with water  ice
resulting in a higher binding energy. A tailored model is necessary to
confirm  the viability  of this  mechanism as  an explanation  for the
outer DCO$^+$ ring.

Finally,  another plausible  explanation is  a local  decrease of  the
ortho  to  para   (o/p)  ratio  of  H$_2$.   The   reaction  shown  in
Eq.\ref{eq:low_T} has a higher activation barrier for p-H$_2$ than for
o-H$_2$  which  will  result  in  an  increase  on  DCO$^+$  formation
\citep{Murillo2015,Walmsley2004}.   Detailed   chemical  modelling  is
needed to test this scenario.

\subsection{Deuterium fractionation}

Our    estimated    disk-averaged    fractionation    ratio    D$_{\rm
f}$(DCO$^+$/HCO$^+$)   of    0.05$\pm$0.01   agrees    with   previous
measurements on  this source  \citep{Huang2017} or even  towards other
disks,   e.g.     DM   Tau    \citep{Guilloteau2006}   and    TW   Hya
\citep{vanDishoeck2003}.

Our estimated value of 0.02$\pm$0.01 for D$_{\rm f}$(DCN/HCN) is lower
than     the     D$_{\rm      f}$(DCO$^+$/HCO$^+$)     and     D$_{\rm
f}$(N$_2$D$^+$/N$_2$H$^+$).   A lower  value of  D$_{\rm
f}$(DCN/HCN) compared  to D$_{\rm f}$(N$_2$D$^+$/N$_2$H$^+$)  could be
hinting at a  more efficient deuterium enrichment in  the outer colder
disk.   On    the   other    hand,   in   comparison    with   D$_{\rm
f}$(DCO$^+$/HCO$^+$), a  lower value  of D$_{\rm f}$(DCN/HCN)  could be
explained  by  the different  deuteration  pathways  that formed  them,
supporting the idea  of a colder formation environment  for DCO$^+$ at
large radii. Both of these values are of the  same  order  of
magnitude as for other protoplanetary disks  \citep{Oberg2012,Huang2017}. The derived D$_{\rm f}$(DCO$^+$/HCO$^+$) value is one order of magnitude lower than in starless cores but similar to the value towards low-mass protostars, such as IRAS 16293-2422 \citep{Butner1995,Caselli2002,Tafalla2006,Schoier2002}. The derived D$_{\rm f}$(DCN/HCN) value is the same order of magnitude towards low-mass protostars and starless cores \citep{Wootten1987,Roberts2002}.

 Our     estimated     value     of    0.45$\pm$0.21     for     D$_{\rm
f}$(N$_2$D$^+$/N$_2$H$^+$) is higher than D$_{\rm f}$(DCO$^+$/HCO$^+$)
and D$_{\rm  f}$(DCO$^+$/HCO$^+$) as predicted by  the chemical models
of \citet{Willacy2007}.  It is also  comparable to the  values towards
the T Tauri disk AS 209 \citep{Huang2015} and consistent with the wide range of ratios, from $<$0.02 to 0.44, found towards starless cores \citep{Crapsi2005}. As noted in the case of AS
209, the  higher D/H ratio  from N$_2$D$^+$  compared to the  one from
DCO$^+$ is  a consequence of  their formation environments.  DCO$^+$ is
formed where CO is not frozen-out  in the relatively warm upper layers
of the  disk. N$_2$D$^+$ is more  abundant where CO is  frozen-out and
deuteration  is enhanced  by Eq.~\ref{eq:low_T}  which, together  with
N$_2$H$^+$, makes it an excellent tracer of the cold outer midplane.

\section{Summary}
\label{sec:con} We have successfully detected three deuterated species
towards HD 163296:  DCO$^+$, N$_2$D$^+$ and DCN. We  use simple models
and estimates of  the midplane temperature to fit  radial rings with
constant abundance  constraining their  relative abundance  and radial
location. Our main conclusions are:

\begin{itemize}

\item We confirm the location of the CO snowline using the second ring
of DCO$^+$ as  a tracer at $\sim$100 AU, consistent  with the previously
reported value  of 90 AU. Our  N$_2$D$^+$ detection is too noisy  to effectively
constrain  its location,  although it  is still  consistent given  our
large beam.

\item DCO$^+$ and DCN show an inner depression that arises most likely
due to the due to
optically thick dust and not due to inherent formation pathway of both species.

\item DCO$^+$ shows a three ring-like structure, located at 70 AU, 150
AU and  260 AU  that could  be linked  to the  structure of  the ${\rm
\mu}$m-sized grains. The two first rings correspond to the high and low
temperature deuteration pathways and are in agreement with simple  DCN and
N$_2$D$^+$ best-fit models.

\item The origin of the third DCO$^+$ ring at 260 AU may be due to a local
decrease  of  UV  opacity  allowing  the  photodesorption  of  CO  and
consequent   formation   of   DCO$^+$   as   seen   in   other   disks
\citep{Oberg2015}, thermal desorption of CO as proposed by \citet{Facchini2017},  or a local decrease of the o/p ratio of H$_2$.

\item The derived  D$_{\rm f}$ values are several  orders of magnitude
higher than the D/H cosmic ratio as expected and are in agreement with
previous measurements and models. The  higher D/H value for N$_2$D$^+$
in  comparison  with those  from  DCO$^+$  and  DCN suggest  a  cooler
formation environment for the former.

\end{itemize}

\begin{acknowledgements}
The authors acknowledge support
by Allegro, the European ALMA Regional Center node in The Netherlands, and
expert advice from Luke Maud in particular. This work
was partially supported by grants from the Netherlands Organization for
Scientific Research (NWO) and the Netherlands Research School for Astronomy
(NOVA) This paper makes use of the following
ALMA data: ADS/JAO.ALMA\# 2013.1.01268.S.
ALMA is a partnership of ESO (representing its member states), NSF (USA)
and NINS (Japan), together with NRC (Canada), NSC and ASIAA (Taiwan),
and KASI (Republic of Korea), in cooperation with the Republic of Chile. The
Joint ALMA Observatory is operated by ESO, AUI/NRAO and NAOJ. 
\end{acknowledgements}

\bibliographystyle{aa}     
\bibliography{bibHD163}

\begin{thebibliography}{55}
\expandafter\ifx\csname natexlab\endcsname\relax\def\natexlab#1{#1}\fi

\bibitem[{{Albertsson} {et~al.}(2013){Albertsson}, {Semenov}, {Vasyunin},
  {Henning}, \& {Herbst}}]{Albertsson2013}
{Albertsson}, T., {Semenov}, D.~A., {Vasyunin}, A.~I., {Henning}, T., \&
  {Herbst}, E. 2013, \apjs, 207, 27

\bibitem[{{Brinch} \& {Hogerheijde}(2010)}]{Brinch2010}
{Brinch}, C. \& {Hogerheijde}, M.~R. 2010, \aap, 523, A25

\bibitem[{{Butner} {et~al.}(1995){Butner}, {Lada}, \& {Loren}}]{Butner1995}
{Butner}, H.~M., {Lada}, E.~A., \& {Loren}, R.~B. 1995, \apj, 448, 207

\bibitem[{{Carney} {et~al.}(2017){Carney}, {Hogerheijde}, {Loomis}, {Salinas},
  {{\"O}berg}, {Qi}, \& {Wilner}}]{Carney2017}
{Carney}, M.~T., {Hogerheijde}, M.~R., {Loomis}, R.~A., {et~al.} 2017, ArXiv
  e-prints 1705.10188]

\bibitem[{{Caselli} {et~al.}(2002){Caselli}, {Walmsley}, {Zucconi}, {Tafalla},
  {Dore}, \& {Myers}}]{Caselli2002}
{Caselli}, P., {Walmsley}, C.~M., {Zucconi}, A., {et~al.} 2002, \apj, 565, 344

\bibitem[{{Ceccarelli} {et~al.}(2007){Ceccarelli}, {Caselli}, {Herbst},
  {Tielens}, \& {Caux}}]{Ceccarelli2007}
{Ceccarelli}, C., {Caselli}, P., {Herbst}, E., {Tielens}, A.~G.~G.~M., \&
  {Caux}, E. 2007, Protostars and Planets V, 47

\bibitem[{{Cleeves}(2016)}]{Cleeves2016}
{Cleeves}, L.~I. 2016, \apjl, 816, L21

\bibitem[{{Crapsi} {et~al.}(2005){Crapsi}, {Caselli}, {Walmsley}, {Myers},
  {Tafalla}, {Lee}, \& {Bourke}}]{Crapsi2005}
{Crapsi}, A., {Caselli}, P., {Walmsley}, C.~M., {et~al.} 2005, \apj, 619, 379

\bibitem[{{Cuppen} \& {Herbst}(2007)}]{Cuppen2007}
{Cuppen}, H.~M. \& {Herbst}, E. 2007, \apj, 668, 294

\bibitem[{{Dalgarno} \& {Lepp}(1984)}]{Dalgarno1984}
{Dalgarno}, A. \& {Lepp}, S. 1984, \apjl, 287, L47

\bibitem[{{Dullemond} \& {Dominik}(2004)}]{Dullemond2004b}
{Dullemond}, C.~P. \& {Dominik}, C. 2004, \aap, 417, 159

\bibitem[{{Dumouchel} {et~al.}(2010){Dumouchel}, {Faure}, \&
  {Lique}}]{Dumouchel2010}
{Dumouchel}, F., {Faure}, A., \& {Lique}, F. 2010, \mnras, 406, 2488

\bibitem[{{Facchini} {et~al.}(2017){Facchini}, {Birnstiel}, {Bruderer}, \& {van
  Dishoeck}}]{Facchini2017}
{Facchini}, S., {Birnstiel}, T., {Bruderer}, S., \& {van Dishoeck}, E.~F. 2017,
  ArXiv e-prints 1705.06235]

\bibitem[{{Favre} {et~al.}(2015){Favre}, {Bergin}, {Cleeves}, {Hersant}, {Qi},
  \& {Aikawa}}]{Favre2015}
{Favre}, C., {Bergin}, E.~A., {Cleeves}, L.~I., {et~al.} 2015, \apjl, 802, L23

\bibitem[{{Flower}(1999)}]{Flower1999}
{Flower}, D.~R. 1999, \mnras, 305, 651

\bibitem[{{Garufi} {et~al.}(2014){Garufi}, {Quanz}, {Schmid}, {Avenhaus},
  {Buenzli}, \& {Wolf}}]{Garufi2014}
{Garufi}, A., {Quanz}, S.~P., {Schmid}, H.~M., {et~al.} 2014, \aap, 568, A40

\bibitem[{{Gerner} {et~al.}(2015){Gerner}, {Shirley}, {Beuther}, {Semenov},
  {Linz}, {Albertsson}, \& {Henning}}]{Gerner2015}
{Gerner}, T., {Shirley}, Y.~L., {Beuther}, H., {et~al.} 2015, \aap, 579, A80

\bibitem[{{Guilloteau} {et~al.}(2006){Guilloteau}, {Pi{\'e}tu}, {Dutrey}, \&
  {Gu{\'e}lin}}]{Guilloteau2006}
{Guilloteau}, S., {Pi{\'e}tu}, V., {Dutrey}, A., \& {Gu{\'e}lin}, M. 2006,
  \aap, 448, L5

\bibitem[{{Huang} \& {{\"O}berg}(2015)}]{Huang2015}
{Huang}, J. \& {{\"O}berg}, K.~I. 2015, \apjl, 809, L26

\bibitem[{{Huang} {et~al.}(2017){Huang}, {{\"O}berg}, {Qi}, {Aikawa},
  {Andrews}, {Furuya}, {Guzm{\'a}n}, {Loomis}, {van Dishoeck}, \&
  {Wilner}}]{Huang2017}
{Huang}, J., {{\"O}berg}, K.~I., {Qi}, C., {et~al.} 2017, \apj, 835, 231

\bibitem[{Isella {et~al.}(2016)Isella, Guidi, Testi, Liu, Li, Li, Weaver,
  Boehler, Carperter, De~Gregorio-Monsalvo, Manara, Natta, P\'erez, Ricci,
  Sargent, Tazzari, \& Turner}]{Isella2016}
Isella, A., Guidi, G., Testi, L., {et~al.} 2016, Phys. Rev. Lett., 117, 251101

\bibitem[{{Mathews} {et~al.}(2013){Mathews}, {Klaassen}, {Juh{\'a}sz},
  {Harsono}, {Chapillon}, {van Dishoeck}, {Espada}, {de Gregorio-Monsalvo},
  {Hales}, {Hogerheijde}, {Mottram}, {Rawlings}, {Takahashi}, \&
  {Testi}}]{Mathews2013}
{Mathews}, G.~S., {Klaassen}, P.~D., {Juh{\'a}sz}, A., {et~al.} 2013, \aap,
  557, A132

\bibitem[{{Millar} {et~al.}(1989){Millar}, {Bennett}, \& {Herbst}}]{Millar1989}
{Millar}, T.~J., {Bennett}, A., \& {Herbst}, E. 1989, \apj, 340, 906

\bibitem[{{Mumma} \& {Charnley}(2011)}]{Mumma2011}
{Mumma}, M.~J. \& {Charnley}, S.~B. 2011, \araa, 49, 471

\bibitem[{{Murillo} {et~al.}(2015){Murillo}, {Bruderer}, {van Dishoeck},
  {Walsh}, {Harsono}, {Lai}, \& {Fuchs}}]{Murillo2015}
{Murillo}, N.~M., {Bruderer}, S., {van Dishoeck}, E.~F., {et~al.} 2015, \aap,
  579, A114

\bibitem[{{{\"O}berg} {et~al.}(2011){{\"O}berg}, {Boogert}, {Pontoppidan}, {van
  den Broek}, {van Dishoeck}, {Bottinelli}, {Blake}, \& {Evans}}]{Oberg2011}
{{\"O}berg}, K.~I., {Boogert}, A.~C.~A., {Pontoppidan}, K.~M., {et~al.} 2011,
  \apj, 740, 109

\bibitem[{{{\"O}berg} {et~al.}(2015){{\"O}berg}, {Furuya}, {Loomis}, {Aikawa},
  {Andrews}, {Qi}, {van Dishoeck}, \& {Wilner}}]{Oberg2015}
{{\"O}berg}, K.~I., {Furuya}, K., {Loomis}, R., {et~al.} 2015, \apj, 810, 112

\bibitem[{{{\"O}berg} {et~al.}(2010){{\"O}berg}, {Qi}, {Fogel}, {Bergin},
  {Andrews}, {Espaillat}, {van Kempen}, {Wilner}, \& {Pascucci}}]{Oberg2010}
{{\"O}berg}, K.~I., {Qi}, C., {Fogel}, J.~K.~J., {et~al.} 2010, \apj, 720, 480

\bibitem[{{{\"O}berg} {et~al.}(2012){{\"O}berg}, {Qi}, {Wilner}, \&
  {Hogerheijde}}]{Oberg2012}
{{\"O}berg}, K.~I., {Qi}, C., {Wilner}, D.~J., \& {Hogerheijde}, M.~R. 2012,
  \apj, 749, 162

\bibitem[{{Perryman} {et~al.}(1997){Perryman}, {Lindegren}, {Kovalevsky},
  {Hoeg}, {Bastian}, {Bernacca}, {Cr{\'e}z{\'e}}, {Donati}, {Grenon},
  {Grewing}, {van Leeuwen}, {van der Marel}, {Mignard}, {Murray}, {Le Poole},
  {Schrijver}, {Turon}, {Arenou}, {Froeschl{\'e}}, \&
  {Petersen}}]{Perryman1997}
{Perryman}, M.~A.~C., {Lindegren}, L., {Kovalevsky}, J., {et~al.} 1997, \aap,
  323, L49

\bibitem[{{Qi} {et~al.}(2011){Qi}, {D'Alessio}, {{\"O}berg}, {Wilner},
  {Hughes}, {Andrews}, \& {Ayala}}]{Qi2011}
{Qi}, C., {D'Alessio}, P., {{\"O}berg}, K.~I., {et~al.} 2011, \apj, 740, 84

\bibitem[{{Qi} {et~al.}(2015){Qi}, {{\"O}berg}, {Andrews}, {Wilner}, {Bergin},
  {Hughes}, {Hogherheijde}, \& {D'Alessio}}]{Qi2015}
{Qi}, C., {{\"O}berg}, K.~I., {Andrews}, S.~M., {et~al.} 2015, \apj, 813, 128

\bibitem[{{Qi} {et~al.}(2013){Qi}, {{\"O}berg}, {Wilner}, {D'Alessio},
  {Bergin}, {Andrews}, {Blake}, {Hogerheijde}, \& {van Dishoeck}}]{Qi2013}
{Qi}, C., {{\"O}berg}, K.~I., {Wilner}, D.~J., {et~al.} 2013, Science, 341, 630

\bibitem[{{Qi} {et~al.}(2008){Qi}, {Wilner}, {Aikawa}, {Blake}, \&
  {Hogerheijde}}]{Qi2008}
{Qi}, C., {Wilner}, D.~J., {Aikawa}, Y., {Blake}, G.~A., \& {Hogerheijde},
  M.~R. 2008, \apj, 681, 1396

\bibitem[{{Remijan} {et~al.}(2003){Remijan}, {Snyder}, {Friedel}, {Liu}, \&
  {Shah}}]{Remijan2003}
{Remijan}, A., {Snyder}, L.~E., {Friedel}, D.~N., {Liu}, S.-Y., \& {Shah},
  R.~Y. 2003, \apj, 590, 314

\bibitem[{{Roberts} {et~al.}(2002){Roberts}, {Fuller}, {Millar}, {Hatchell}, \&
  {Buckle}}]{Roberts2002}
{Roberts}, H., {Fuller}, G.~A., {Millar}, T.~J., {Hatchell}, J., \& {Buckle},
  J.~V. 2002, \aap, 381, 1026

\bibitem[{{Sch{\"o}ier} {et~al.}(2002){Sch{\"o}ier}, {J{\o}rgensen}, {van
  Dishoeck}, \& {Blake}}]{Schoier2002}
{Sch{\"o}ier}, F.~L., {J{\o}rgensen}, J.~K., {van Dishoeck}, E.~F., \& {Blake},
  G.~A. 2002, \aap, 390, 1001

\bibitem[{{Sch{\"o}ier} {et~al.}(2005){Sch{\"o}ier}, {van der Tak}, {van
  Dishoeck}, \& {Black}}]{LAMBDA_database2005}
{Sch{\"o}ier}, F.~L., {van der Tak}, F.~F.~S., {van Dishoeck}, E.~F., \&
  {Black}, J.~H. 2005, \aap, 432, 369

\bibitem[{{Tafalla} {et~al.}(2006){Tafalla}, {Santiago-Garc{\'{\i}}a}, {Myers},
  {Caselli}, {Walmsley}, \& {Crapsi}}]{Tafalla2006}
{Tafalla}, M., {Santiago-Garc{\'{\i}}a}, J., {Myers}, P.~C., {et~al.} 2006,
  \aap, 455, 577

\bibitem[{{Teague} {et~al.}(2015){Teague}, {Semenov}, {Guilloteau}, {Henning},
  {Dutrey}, {Wakelam}, {Chapillon}, \& {Pietu}}]{Teague2015}
{Teague}, R., {Semenov}, D., {Guilloteau}, S., {et~al.} 2015, \aap, 574, A137

\bibitem[{{Turner}(2001)}]{Turner2001}
{Turner}, B.~E. 2001, \apjs, 136, 579

\bibitem[{{van den Ancker} {et~al.}(1997){van den Ancker}, {The}, {Tjin A
  Djie}, {Catala}, {de Winter}, {Blondel}, \& {Waters}}]{vandenAncker1997}
{van den Ancker}, M.~E., {The}, P.~S., {Tjin A Djie}, H.~R.~E., {et~al.} 1997,
  \aap, 324, L33

\bibitem[{{van Dishoeck} {et~al.}(2003){van Dishoeck}, {Thi}, \& {van
  Zadelhoff}}]{vanDishoeck2003}
{van Dishoeck}, E.~F., {Thi}, W.-F., \& {van Zadelhoff}, G.-J. 2003, \aap, 400,
  L1

\bibitem[{{van't Hoff} {et~al.}(2017){van't Hoff}, {Walsh}, {Kama}, {Facchini},
  \& {van Dishoeck}}]{vantHoff2016}
{van't Hoff}, M.~L.~R., {Walsh}, C., {Kama}, M., {Facchini}, S., \& {van
  Dishoeck}, E.~F. 2017, \aap, 599, A101

\bibitem[{{Vidal-Madjar}(1991)}]{Vidal-Madjar1991}
{Vidal-Madjar}, A. 1991, Advances in Space Research, 11, 97

\bibitem[{{Walmsley} {et~al.}(2004){Walmsley}, {Flower}, \& {Pineau des
  For{\^e}ts}}]{Walmsley2004}
{Walmsley}, C.~M., {Flower}, D.~R., \& {Pineau des For{\^e}ts}, G. 2004, \aap,
  418, 1035

\bibitem[{{Watson}(1976)}]{Watson1976}
{Watson}, W.~D. 1976, Reviews of Modern Physics, 48, 513

\bibitem[{{Willacy}(2007)}]{Willacy2007}
{Willacy}, K. 2007, \apj, 660, 441

\bibitem[{{Wilson}(1999)}]{Wilson1999}
{Wilson}, T.~L. 1999, Reports on Progress in Physics, 62, 143

\bibitem[{{Wisniewski} {et~al.}(2008){Wisniewski}, {Clampin}, {Grady},
  {Ardila}, {Ford}, {Golimowski}, {Illingworth}, \& {Krist}}]{Wisniewski2008}
{Wisniewski}, J.~P., {Clampin}, M., {Grady}, C.~A., {et~al.} 2008, \apj, 682,
  548

\bibitem[{{Woods} \& {Willacy}(2009)}]{Woods2009}
{Woods}, P.~M. \& {Willacy}, K. 2009, \apj, 693, 1360

\bibitem[{{Wootten}(1987)}]{Wootten1987}
{Wootten}, A. 1987, in IAU Symposium, Vol. 120, Astrochemistry, ed. M.~S.
  {Vardya} \& S.~P. {Tarafdar}, 311--318

\bibitem[{{Yen} {et~al.}(2016){Yen}, {Koch}, {Liu}, {Puspitaningrum}, {Hirano},
  {Lee}, \& {Takakuwa}}]{Yen2016}
{Yen}, H.-W., {Koch}, P.~M., {Liu}, H.~B., {et~al.} 2016, \apj, 832, 204

\bibitem[{{Zhang} {et~al.}(2016){Zhang}, {Bergin}, {Blake}, {Cleeves},
  {Hogerheijde}, {Salinas}, \& {Schwarz}}]{Zhang2016}
{Zhang}, K., {Bergin}, E.~A., {Blake}, G.~A., {et~al.} 2016, \apjl, 818, L16

\bibitem[{{Zhang} {et~al.}(2017){Zhang}, {Bergin}, {Blake}, {Cleeves}, \&
  {Schwarz}}]{Zhang2017}
{Zhang}, K., {Bergin}, E.~A., {Blake}, G.~A., {Cleeves}, L.~I., \& {Schwarz},
  K.~R. 2017, Nature Astronomy, 1, 0130

\end{thebibliography}

\begin{appendix}
\section{Keplerian Masking}
\label{appendix}
\begin{figure*}[!h]
\centering
\begin{subfigure}{1.0\textwidth}
        \centering
        \includegraphics[width=1.0\textwidth]{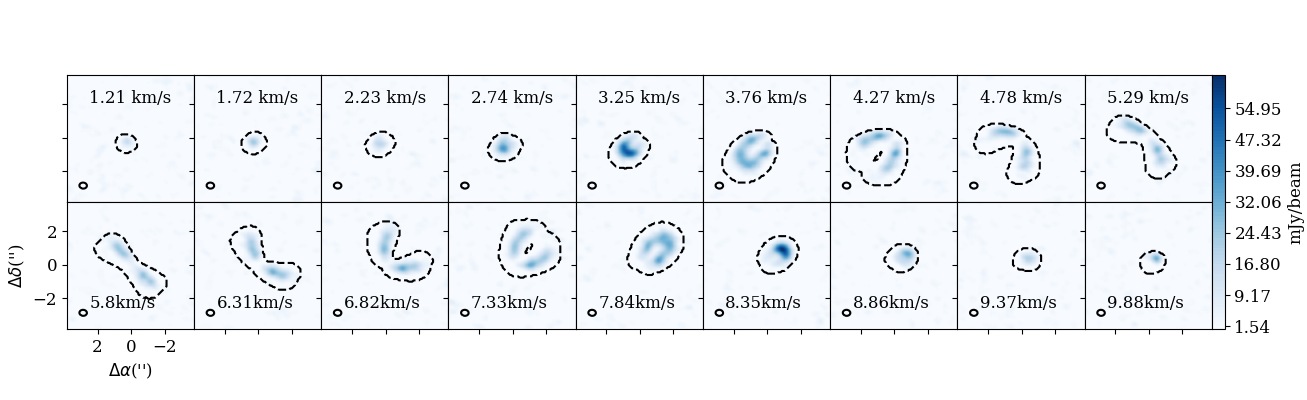}
\end{subfigure}
\begin{subfigure}{1.0\textwidth}
        \centering
        \includegraphics[width=1.0\textwidth]{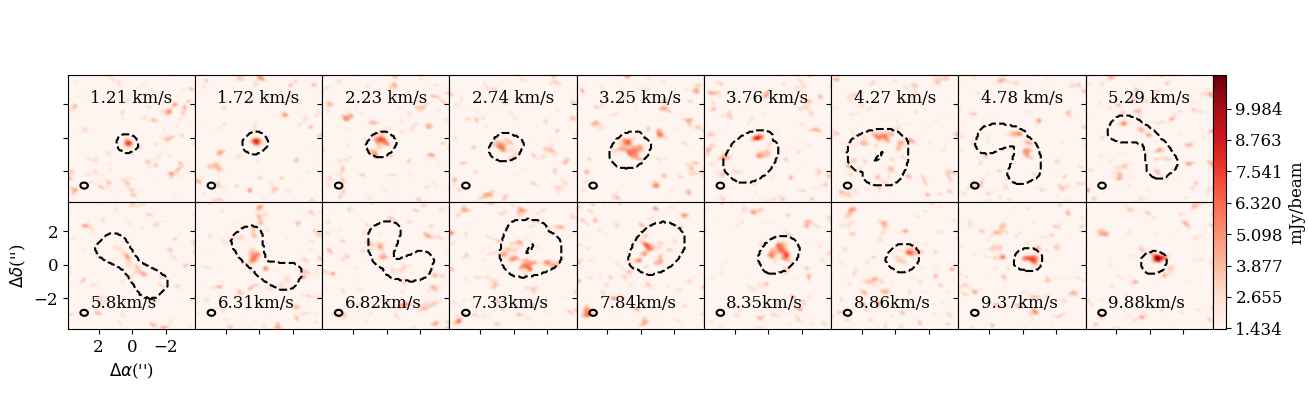}
\end{subfigure}
\begin{subfigure}{1.0\textwidth}		
        \centering
        \includegraphics[width=1.0\textwidth]{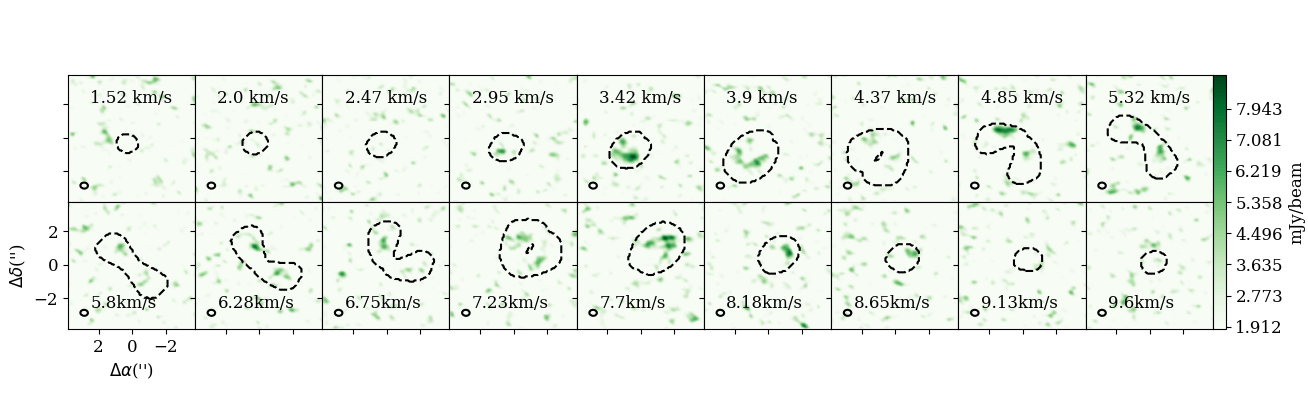}        
\end{subfigure} 

\caption{Overplotted Keplerian mask, in black-dashed contours, and spectral maps of DCO$^+$  $J$=3--2, DCN
$J$=3--2 and N$_2$D$^+$  $J$=3--2.}
\label{fig:masking}

\end{figure*}

Since the emission  in each channel will only come from
a region in the sky where  the rotational velocity of the disk matches
the Doppler  shift of the line, selecting a subset  of pixels on
each velocity channel with Keplerian  velocities equal to the expected
Doppler shift  enhances the  signal to noise  of the  final integrated
intensity  map. To  achieve  this, we  first  calculate the  projected
radial  velocity on  each  pixel using  a stellar  mass  value of  2.3
M$_\odot$ \citep{vandenAncker1997}. We create a mask cube with the same dimensions as the data spectral cube and select, for each spectral plane, the  pixels that  have a  projected radial  velocity greater  than the
correspondent Doppler-shifted velocity v$_r$ of the spectral plane and
lower  than  v$_r$+$\delta  v$,  where  $\delta  v$  is  the  spectral
resolution  of the  cube. Finally, each spectral plane of the mask cube is convolved  with  the
synthesized  beam to  create a final mask that accounts for the smearing of the disk as seen  in
Fig~\ref{fig:masking}.

 A similar approach has already been used by \citet{Yen2016}, stacking
azimuthal regions that are corrected by their Keplerian velocity. They
reconstruct integrated  intensity maps by Doppler-shifting  a selected
azimuthal region within an inner  and outer radius. Since both methods
are flux conserving, the only difference lies in reduced noise for
the latter  in sacrifice of  spatial information. The stacking method by \citet{Yen2016} achieves a  S/N increase  of a factor of $\sim$ 4 for the weak lines N$_2$D$^+$ J=3--2 and DCN J=3--2. In contrast our method achieves an increase in S/N of a factor of $\sim$ 2 but results in 
a     much      clearer     integrated     intensity      map     (see
Fig.~\ref{fig:mom_all_lines}).

\end{appendix}

%
%


\end{document}